\documentclass[aps,preprint,groupedaddress,floatfix,showpacs]{revtex4}
\usepackage{amsmath,amssymb,graphicx,color}
\usepackage{bm}

\def\bmu{\mbox{\boldmath{$\mu$}}}
\begin{document}
\title{Nonequilibrium work statistics of an Aharonov-Bohm flux}

\author{Juyeon Yi}

\affiliation{Department of Physics, Pusan National University,
Busan 609-735, Korea}

\author{Peter Talkner, Michele Campisi}
\affiliation{Institute of Physics, University of Augsburg,
Universit\"{a}tsstrasse 1, D-86135, Augsburg, Germany}

\date{\today}

\begin{abstract}
We investigate the statistics of work performed on a noninteracting
electron gas confined into a ring as a threaded magnetic field is
turned on.  For an electron gas initially prepared in a grand canonical state
it is demonstrated that the Jarzynski equality
continues to hold in this case, with the free energy replaced by the
grand potential. The work distribution displays a marked
dependence on the temperature. While in the classical~(high
temperature) regime, the work probability density function follows
a Gaussian distribution and the free energy difference entering
the Jarzynski equality is null, the free energy difference is
finite in the quantum regime, and the work probability distribution
function becomes multimodal. We point out the
dependence of the work statistics on the number of electrons
composing the system.

\end{abstract}


\maketitle
\section{Introduction}

One of the most fascinating electro-magnetic field effects is the modulation of quantum interference in multiply connected spatial regions due to electromagnetic fields.
The Aharonov-Bohm~(AB) effect is a well-known example where a
localized magnetic field introduces the phase shift of a particle
wavefunction and results in an interference pattern governed by
the AB flux $\Phi_{AB}=\oint {\bf A}\cdot d{\bf l}$~\cite{ab}.
A similar effect, called the Aharonov-Casher~(AC)
effect occurs, when a neutral particle with a magnetic moment $
\bmu$
moves in an electric field ${\bf E}$ and
acquires a phase shift amounting to the flux
$\Phi_{AC}=\oint {\bf E}\times d{\bf l}\cdot \bmu /e$
~\cite{ac,ac2,ac3}.
A dual effect was also pointed out for a neutral
particle carrying an electric dipole moment moving in a magnetic
field of the appropriate configuration~\cite{dualac}. In the
mentioned examples, electromagnetic fields influence the wave function and also the energy spectrum of a particle moving in a multiply connected spatial region but do not exert any classical force.

Recently, scientists working in the field of non-equilibrium thermodynamics have
drawn the attention to the fact that the work done by external forces
on a driven system may be usefully employed to characterize their response properties.
Jarzynski introduced the celebrated nonequilibrium work relation that links the free energy difference~($\Delta F$)
to the averaged exponentiated negative work~\cite{Jarzynski,Jarzynski2}:
\begin{equation}\label{JE}
\langle e^{-\beta w}\rangle = e^{-\beta \Delta F},
\end{equation}
where $w$ is the work performed on a system by a
time-dependent force determined by a prescribed protocol and $\langle \cdots \rangle$ denotes the
average over many realizations of the forcing experiment. The
equality was primarily derived for classical systems, which
experiments and theories so far mainly refer to
~\cite{classical1,classical2,classical3}. Fluctuation theorems in presence of magnetic fields and other non-conservative forces were studied for classical systems in Ref.~\cite{Pradhan}.
On the other hand, generalizations of Eq.~(1) to quantum
mechanical systems were
discussed~\cite{quantum1,quantum2,quantum2b,quantum3,TLH,TH,THM,DL,ftopen,Talkner08PRE78,Talkner09PRE79}, for recent reviews see \cite{EHM,CHT}.
In quantum mechanics, the work is obtained by means of two energy measurements
at the beginning and at
the end of a given protocol. In the mentioned examples of quantum interference effects, the electromagnetic fields do work on a charged particle, or a magnetic or electric dipole in multiply connected domains not only caused by the classical forces exerted on the particle but also by the shifts of the energy spectrum~\cite{pc,pc2,pc3}.

The main purpose of this work is to investigate the statistics of the work done by an
Aharonov-Bohm flux for quantum charged particles moving
along a one-dimensional ring representing the simplest possible multiply connected domain. We consider not only the
single-particle case but also many-particle systems by
generalizing the fluctuation theorem for a grand canonical initial
state. Especially we focus on fermionic systems in a ring-configuration (see Fig.~1).
Their equilibrium properties were discussed in terms of persistent currents decades ago~\cite{pc,pc2,pc3,loss}.
Recent measurement using nano-cantilevers to detect changes in the magnetic field produced by the current
has achieved high accuracy and prompted
a renewed interest in this topic~\cite{pcexp}.
It is noteworthy that the focus of our study is laid upon the nonequilibrium nature
of the system, revealed in the statistics of work done by the magnetic flux. By obtaining an analytic expression of the
characteristic function of work, we examine the quantum and
classical nature of the resulting distributions and study their
dependence on both temperature and particle number.

The paper is organized as follows: Sec. II is devoted to the
introduction of the system of interest. In Sec.III, we obtain the
probability distribution for a single-particle case. We then
consider a grand canonical initial state of many-particle systems,
and present the results in Sec.IV. Summary and conclusion are
drawn in Sec.V.

\begin{figure}[b]
\resizebox{10cm}{!}{\includegraphics{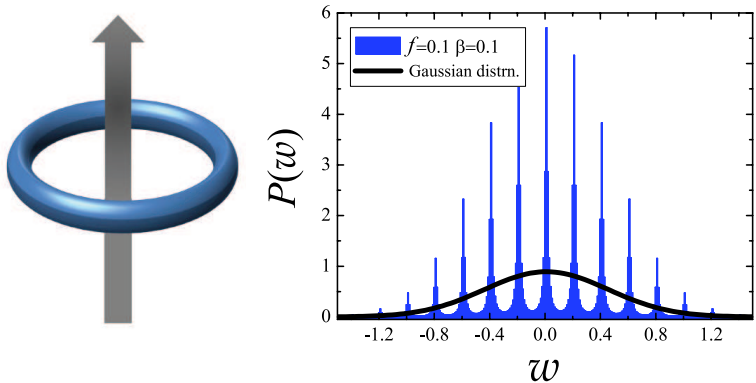}}
 \caption{ \label{fig1}
(Color online)~ The left panel gives a schematic picture of a ring
with a threading flux, and the right panel displays the work
distribution consisting of a series of peaks for a single particle moving
along the ring upon a sudden switch of the dimensionless flux $f=\Phi/\Phi_0$ from $f=0$ to $f=0.1$. The thick solid line represents the Gaussian approximation of the work distribution, Eq.~(\ref{Gauss}), for the same parameter values.  }
\end{figure}

\section{The system}
We consider $N$ spinless fermions
moving along an infinitely thin ring of radius $R$ in presence of a magnetic flux.
The corresponding Hamiltonian reads~\cite{loss}
\begin{equation}\label{firstham}
{\cal H}_f=\frac{\hbar^{2}}{2mR^{2}}\sum_{\ell=1}^{N}\left(\frac{\partial}{i\partial \theta_{\ell}}-f \right)^{2},
\end{equation}
where $\theta_{\ell}$ is the angular coordinate of the $\ell$th particle, and $f=\Phi/\Phi_{0}$ is the total flux threading the ring,
$\Phi$, in units of the flux quantum, $\Phi_{0}=hc/e$.
The single particle energy eigenvalues are given by~\cite{loss}
\begin{equation}
E_{k}(f)=E_{0}(k-f)^{2},
\label{Ef}
\end{equation}
where $E_{0}=\hbar^{2}/2mR^{2}$ characterizes the energy-level
spacing, and the integer $k$ denotes the angular momentum quantum number.

For later consideration of a many-particle system, let us introduce the second
quantized form of the Hamiltonian:
\begin{equation}\label{ham}
{\cal H}_f=\sum_{k}E_{k}(f)c_{k}^{\dagger}c_{k}\equiv
\sum_{k}E_{k}(f){\cal N}_k,
\end{equation}
where $c_{k}^{\dagger}~(c_{k})$ is the creation~(annihilation) operator
of an electron in the $k$th angular
momentum~(or energy) eigenstate, and the number operator
${\cal N}_{k}=c_{k}^{\dagger}c_{k}$ measures the particle
number in the $k$-th state.

We will study the probability distribution function $P(w)$ of the work  $w$ that
is performed on the electrons in the time span $[0,\tau]$, by a magnetic flux $f(t)$ that varies in time. As a consequence, the Hamiltonian of the system becomes time-dependent. It will be denoted by ${\cal H}(t) \equiv {\cal H}_{f(t)}$.
Here we restrict ourselves to the case of a sudden switch of the magnetic flux
immediately after the time $t=0$ with
\begin{equation}
f(t) = \left \{
\begin{array}{ll}
0 \quad &\text{for}\; t\leq 0\\
f&\text{for}\; t > 0
\end{array}
\right .\:.
\label{ft}
\end{equation}
Given this protocol, we will first calculate the work characteristic
function (i.e., the Fourier transform of $P(w)$), which for a canonical initial state is given by the
formula~\cite{TLH,TH,THM,DL}:
\begin{equation}\label{singlecha}
G(u)=\int_{\infty}^{-\infty}dw e^{iuw}P(w)=\langle
e^{iu{\cal H}_{H}(\tau)}e^{-iu{\cal H}(0)}\rangle _{\rho_{c}},
\end{equation}
and then obtain $P(w)$ by inverse Fourier transformation.
Here $\langle X \rangle_{\rho_{c}}=\mbox{Tr}X\rho_{c}$ and $\rho_{c}=e^{-\beta {\cal H}(0)}/{\cal Z}_{0}$
with the normalization ${\cal Z}_{0}$ being the canonical
partition function. Further, ${\cal H}_{H}(t)$ denotes the Hamiltonian operator in the Heisenberg representation.
Note that since $[{\cal H}(t), {\cal H}(s)]=0$ for any $t$ and $s$ in the time span,
the Hamiltonians in the  Schr\"{o}dinger and Heisenberg 
picture coincide, i.e. ${\cal H}_{H}(\tau)={\cal H}(\tau)$.

\section{Single particle case}
In the case of a single particle
system, we obtain from Eqs.~(\ref{firstham}) and (\ref{singlecha}),
\begin{equation}\label{singlecharac}
G(u) = e^{iuE_{0}f^{2}}\sum_{k}e^{2ikf E_{0}u}e^{-\beta
E_{0}k^{2}}/{\cal Z}_{0}.
\end{equation}
We rescale all variables with the dimension of an energy by the
natural energy unit $E_{0}$ as ${\widetilde u}=E_{0}u$,
${\widetilde w}=w/E_{0}$, and ${\widetilde \beta}=\beta
E_{0}$~(for notational simplicity, we drop the tilde in the following). Note that at
high temperatures the system enters the classical regime. In this regime we can
discard the discretness of $k$ and replace the summation by an integration, namely,
$\sum_{k}\rightarrow (R/\hbar)\int_{-\infty}^{\infty}dp$ with the momentum
defined by $p=\hbar k/R$. We thus obtain $G(u) \approx G_{c}(u)=
e^{if^{2}u-f^{2}u^{2}/\beta}$, which leads to a 
Gaussian distribution of work reading
\begin{equation}
P_{c}(w)=\sqrt{\beta/(4\pi
f^{2})}e^{-\beta(w-f^{2})^{2}/(4f^{2})}\:.
\label{Gauss}
\end{equation}
Note that for $u=i\beta$,
$G_{c}(i\beta)=\langle e^{-\beta w}\rangle =1$.
This conversion from discrete summation to integration
is accurate only at
sufficiently high temperatures~($\beta \ll 1$), or for a ring of sufficiently large
radius. When we fully count the level discretness, the work
distribution is expected to be a series of peaks, and the normal
distribution would provide its envelop. This can be
confirmed by evaluating $P_{c}(w)$ directly from
Eq.~(\ref{singlecharac}) via an inverse Fourier transform:
\begin{equation}
P(w)=\sum_{k}{\cal W}_{k}\delta (w-2kf-f^{2})
\end{equation}
where ${\cal W}_{k}=e^{-\beta k^{2}}/\sum_k e^{- \beta k^2}$ is the weight of the $k$th peak. The right panel of
Fig.~1 displays the resulting distribution~\cite{note}.
It is then a due course to
investigate many-particle cases, where the effect of finite
particle number comes into question.

\section{Grand canonical initial state}
  In order to deal with many-particle systems, we extend the characteristic function of work to initial grand canonical states. Although not needed here, we allow
for possible changes of particle numbers. This will lead to a  generalization of the
Jarzynski work theorem to grand canonical initial states.
Only later we will specify to the case of strict particle number conservation.

In the grand canonical ensemble energy and particle number are fluctuating quantities. In order to determine their changes effected by a protocol a simultaneous measurement of energy and particle number must be performed at the beginning and at the end of the protocol, see Ref. \cite{TCH} where the joint statistics of changes of two observables are discussed.
Joint measurements necessitate that the particle number operator $\cal{N}$ commutes with both the initial and final Hamiltonian, i.e., $[{\cal N},{\cal H}(\tau)]=[{\cal N},{\cal H}(0)]=0$.
Then joint eigenfunctions $|\Psi_{\nu,N}(0) \rangle$ and $|\Psi_{{\bar \nu},\bar{N}}(\tau) \rangle$ exist with corresponding pairs of eigenvalues $E_{\nu}$, $N$ and $E_{{\bar{\nu}}}$, $\bar{N}$, satisfying
${\cal H}(\tau)|\Psi_{{\bar \nu},\bar{N}}(\tau)\rangle=E_{{\bar \nu},\bar{N}}(\tau)|\Psi_{{\bar \nu},\bar{N}}(\tau)\rangle$
and ${\cal N}|\Psi_{{\bar \nu}.\bar{N}}(\tau)\rangle ={\bar N}|\Psi_{{\bar \nu},\bar{N}}(\tau)\rangle$, as well as
${\cal H}(0)|\Psi_{\nu,N}(0)\rangle=E_{\nu}(0)|\Psi_{\nu,N}(0)\rangle$
and ${\cal N}|\Psi_{\nu,N}(0)\rangle =N|\Psi_{\nu,N}(0)\rangle$.
The joint probability density function ${\cal P}(w,n)$ to observe the work $w$ and particle number change $n$ in a single realization of the protocol is given by
\begin{eqnarray}\label{subprob}
{\cal
P}(w,n)=\sum_{\nu, {\bar \nu}}\sum_{N, {\bar N}}&\delta(w&-E_{{\bar \nu}}(\tau)+E_{\nu}(0))
\delta_{n,{\bar N}-N}
\\ \nonumber &\times
&P(E_{{\bar \nu}}(\tau),{\bar N}|E_{\nu}(0),N)P^{(eq)}_{g}(E_\nu(0),N) ,
\end{eqnarray}
where $P^{(eq)}_{g}=e^{-\beta(E_{\nu}-\mu N)}/{\cal Q}_{0}$ is the joint probability of finding  the energy $E_\nu(0)$ and particle number $N$ in the initial grand-canonical state; further, ${\cal
Q}_{0}=\sum_{\epsilon_{\nu},N}e^{-\beta(E_{\nu}(0)-\mu
N)}$ that is the grand-canonical partition function.
The conditional probability $P(E_{{\bar \nu}}(\tau),{\bar N}|E_{\nu}(0),N)$ for finding the energy $E_{{\bar \nu}}(\tau)$ and particle number $\bar{N}$ at the end of the protocol once it was $E_\nu(0)$ and $N$ at the beginning 
is determined by the overlap between
the final state and the time evolved initial state:
\begin{equation}\label{jprob}
P(E_{{\bar \nu}}(\tau),{\bar N}|E_{\nu}(0),N)
=|\langle\Psi_{{\bar \nu}}(\tau)|U(\tau,0)|\Psi_{\nu}(0)\rangle |^{2},
\end{equation}
where $U(t,0)$ is the unitary time evolution operator solving the Schr\"odinger equation $i \hbar \partial  U(t,0) /\partial t = {\cal H}(t) U(t,0)$ with $U(0,0) = 1$.
The Fourier transform of the joint probability (\ref{subprob})
then yields the characteristic function that can be cast into the form of a two-time correlation function \cite{TCH}, i.e.,
\begin{eqnarray}\label{grandcha}
{\cal G}(u,v) &=& \sum_{n=-\infty}^\infty \int dw\: e^{iuw+ivn} {\cal P}(w,n) \\
\nonumber
&=&\langle
e^{iu {\cal H}_{H}(\tau)+iv{\widehat {\cal N}}_{{\cal H}}(\tau)}e^{-iu {\cal H}(0)-iv{\widehat {\cal N}}(0)}\rangle_{\rho_{g}} .
\end{eqnarray}
Putting $w=i \beta$ and $v =-i \beta \mu$ one obtains a generalized Jarzynski equality for the grand-canonical initial state reading
\begin{equation} \label{gcje}
\langle e^{-\beta w}e^{-\beta \mu n}\rangle \equiv
\sum_{n=-\infty}^{\infty}\int dw e^{-\beta w}e^{-\beta \mu n}{\cal P}(w,n)=\frac{{\cal Q}_{\tau}}{{\cal Q}_{0}}
=e^{-\beta \Delta \Omega}
\end{equation}
with the grand potential difference $\Delta \Omega = \Omega(\tau)-\Omega(0)$ where $\Omega(t)= -\beta^{-1} \ln {\cal Q}(t)$.
Similar considerations were made for  classical
systems~\cite{seifert} and also for composed quantum systems with number
exchanges between subsystems~\cite{SU,qmnumber,CTH}.

\section{many-particle case}
We analyze the work statistics of a many electron system undergoing a sudden switch of the magnetic flux by means of the generalized  Eq.~(\ref{gcje}).
Since in our case
the particle number is a constant of motion,
${\cal N}_{H}(\tau)={\cal N}(0)$, the characteristic function, Eq.~(\ref{grandcha}), is independent of
$v$;
therefore we simply write it as ${\cal G}(u,v) \equiv {\cal G}(u)$.
Due to the sudden switch of the magnetic flux
${\cal H}_{H}(\tau)={\cal H}(\tau)=\sum_{k}E_{k}(f){\cal N}_{k}$ as given by Eq.~(3).
Moreover ${\cal H}(0)$ and ${\cal H}(\tau)$ commute with each other,
and we can then write
\begin{equation}
{\cal G}(u)=\langle e^{iu\sum_{k}\Delta_{k}(f){\cal N}_{k}}\rangle_{\rho_{g}}
=\prod_{k}[1-\langle {\cal N}_{k}\rangle +\langle
{\cal N}_{k}\rangle e^{iu\Delta_{k}(f)}],
\end{equation}
where we used the property ${\cal N}_{k}^{2}={\cal N}_{k}$ of fermionic number operators.
Here $\Delta_{k}(f)=E_{k}(f)-E_{k}(0)$ and $\langle {\cal N}_{k}\rangle
=[1+e^{\beta(E_{k}(0)-\mu)}]^{-1}$ for the fermionic particles.
\begin{figure}
\resizebox{10cm}{!}{\includegraphics{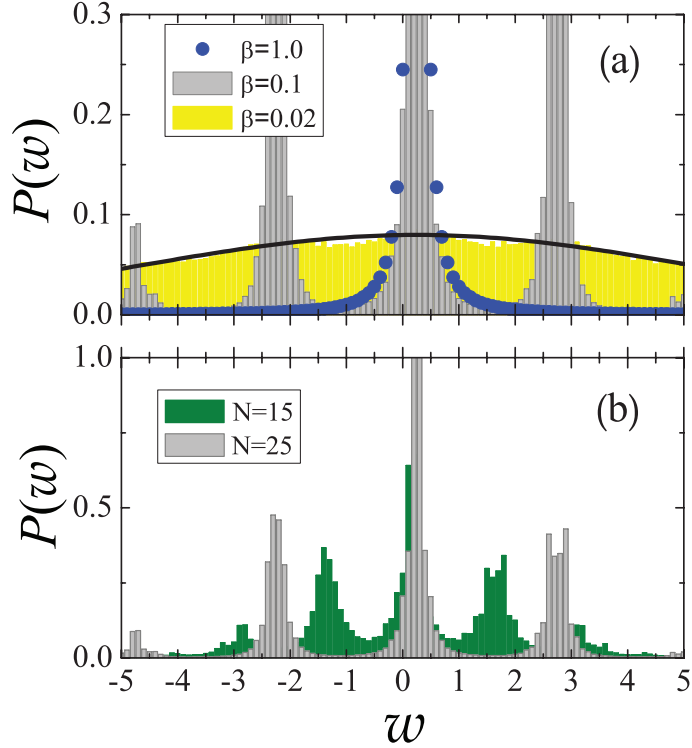}}
 \caption{
(Color online)~ The probability distribution ${\cal P}(w)$ of the collective
work performed on $N$ spinless fermions by a sudden switch of the magnetic flux from $f=0$ to $f=0.1$.
In panel (a) results are displayed  for the average number of particles  $N=25$ at three different temperatures.
The solid line represents the high temperature Gaussian approximation Eq. (\ref{Gauss}) for $\beta=0.02$.
Panel (b) displays the distributions for two different average particle numbers at inverse temperature $\beta=0.1$. Note that for the larger $N$ value the side peaks have a larger distance from the central peak.}
\end{figure}
The chemical potential $\mu$ should be determined to satisfy
\begin{equation}
\langle {\cal N}\rangle =\sum_{k}\frac{1}{1+e^{\beta(E_{k}-\mu)}},
\label{Nmu}
\end{equation}
where $\langle {\cal N}\rangle$ is the average number of particles, which will be denoted as
$N$ hereafter.

Figure~2
shows the work distributions for the flux $f=0.1$, at
different temperatures and particle-numbers. As shown in Fig.~2(a)
at low temperatures~(large $\beta$), the distribution is narrow and centered at
the difference between the ground-state energies in presence and absence of the magnetic flux which is given by $w_{c}=\sum_{k\in k_{gs}}(f^{2}-2kf)$. Here the summation runs
over the $k$-values given by $k_{gs}=0,\pm 1, \cdots, \pm(N-1)/2$ for odd $N$.
Due to the pairwise cancelation of positive and negative $k$-values in
$w_{c}$, the term linear in $f$ vanishes, and hence
$w_{c}=Nf^{2}$. For the used parameters $N=25$ and $f=0.1$,
$w_{c}=0.25$ which indeed coincides with the central peak
positions in Fig.~2(a). 
At higher temperatures~(see $\beta =0.1$ in Fig.~2a),
excited states of $k=\pm (N+1)/2$ come into play, which lead to
side peaks located at $w_{s}=w_{c}\pm (N+1)f$. The number
dependence of the side peak positions can be seen in Fig.~2(b):
With decreasing particle numbers the distances between the central peak and the side peaks shrink.
At
high-temperatures, many excited levels contribute to the work
fluctuation with almost equal weights. This leads to the seemingly
continuous and flat distribution as displayed for $\beta=0.02$ in
Fig.~2(a). In this case, in fact, particles follow the
Maxwell-Boltzmann statistics, and the energy spectrum can be
conceived as a continuum. Then, the
characteristic function is approximately given by the products of
$N$ single particle contributions, i.e., by ${\cal
G}_{c}(u)\approx G_{c}^{N}(u)$. This gives ${\cal P}_{c}(w)\approx
\sqrt{\beta/(4\pi N f^{2})}e^{-\beta(w-Nf^{2})^{2}/(4Nf^{2})}$
which is plotted as solid line for $\beta=0.02$ in Fig.~2(a).
\begin{figure}[t]
\resizebox{10cm}{!}{\includegraphics{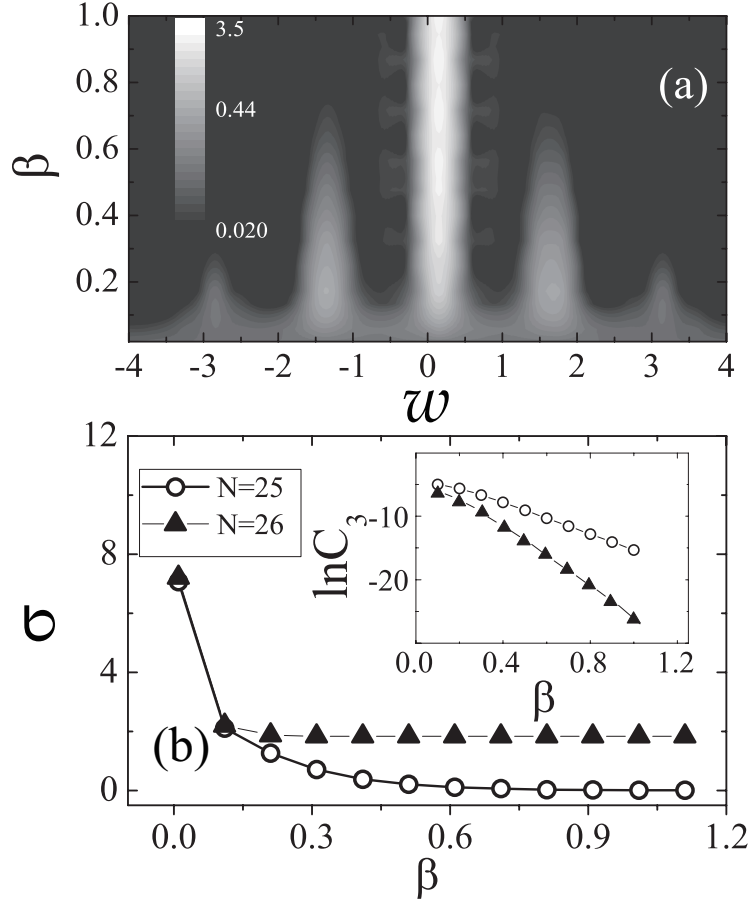}}
 \caption{ \label{fig2}
(Color online)~ (a) The probability distribution of work $w$ for $N=15$ under a sudden magnetization flux switch from $f=0$ to $f=0.1$ is displayed as a function of  $w$ and the inverse temperature $\beta$ by means of different gray-values as specified in the left upper part of the panel. At low temperatures the distribution is unimodal and develops side peaks with increasing temperature.
(b) The temperature
dependence of the standard deviation of work $\sigma$ is compared for $N=25$ and
$N=26$. In the case of the even average particle number the standard deviation saturates at a finite value with decreasing temperature while it vanishes for the odd average particle number.
The inset shows the natural logarithm of the third
cummulant~($C_{3}$) versus $\beta$. }
\end{figure}
We present the temperature dependence of the probability distribution
in Fig.~3(a). It displays the change from a narrow unimodal distribution at low temperatures through a multiply peaked distribution at intermediate temperatures to a broad Gaussian distribution at high temperatures.
From the characteristic function of work
one obtains the variance and all $n$th-order cummulants, $C_{n}$ via the formula,
$C_{n}=(-i)^{n}\partial_{u}^{n}\ln {\cal
G}(u)|_{u=0}$, where $\partial_{u}^{n}$ denotes the $n$th
derivative with respect to $u$. As shown in Fig.~3(b) the variance
and the third order cummulant for $N=25$ rapidly decrease to zero
with
decreasing temperature. The inset shows the exponential
temperature dependence of $C_{3}$. We note that also the variance $\sigma^2= C_{2}$ decays exponentially with temperature, although we do not shown it here. On
the other hand, for $N=26$ the variance saturates to a finite
value, while the third moment exhibits an exponential decay,
similarly to, but far faster than for the case when $N=25$.

\begin{figure}
\resizebox{10cm}{!}{\includegraphics{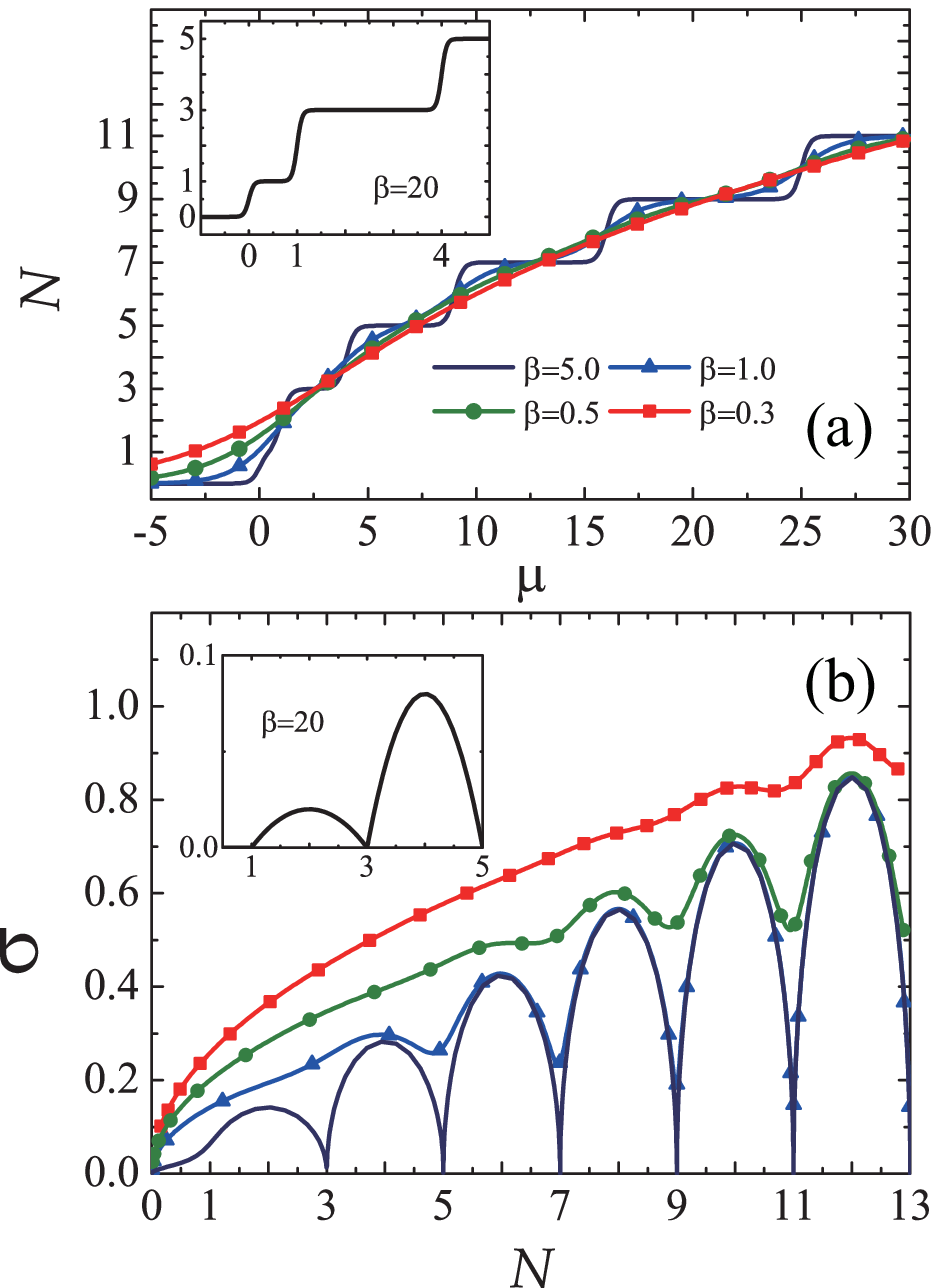}}
\caption{
(Color online)~ (a) The average number of particles as a function of the chemical potential for various temperatures. The stepwise increase of $N$ can be seen at low temperatures (see the curve for $\beta =5.0$ and the inset for $\beta=20$).
(b) The number dependence of the standard deviation of the work at $f=0.1$. At the low temperatures  $\sigma$are displays narrow dips at odd $N$ and broad peaks at even $N$. At the lowest temperature $\beta=20$ shown in the inset the standard deviation is vanishingly small up to $N=1$ and then displays a maximum at $N=2$, see the inset. At high temperatures ($\beta =0.3$), those structures are washed out at small and intermediate $N$ and become visible only for sufficiently large $N$.
}
\end{figure}
The explicit form of the variance for the system
is given by
\begin{equation}
\sigma^{2}= \sum_{k}[ (1-\langle {\cal N}_{k}\rangle)\langle
{\cal N}_{k}\rangle\Delta^{2}_{k}(f)]
\label{var}
\end{equation}
which indicates that the number fluctuations $(1-\langle {\cal N}_k \rangle )\langle {\cal N}_k \rangle$ determine the
variance of the work.
The total average number $N$ of particles in the initial equilibrium state in absence of a magnetic flux is given by Eq.~(\ref{Nmu}). At low temperatures it increases in a
stepwise fashion with varying $\mu$ and forms plateaux of height $N=2n+1$
with steps close to $\mu = n^2$~(see Fig.~4(a)).
This behavior is a direct consequence of the degeneracy of states with angular momentum $\pm k$.
Due to the jumps, a system with an
even average number of particles has pronounced work fluctuations that persist with decreasing temperature whereas for an odd number the variance of work vanishes with decreasing temperature.
We present the number dependence of the standard deviation in panel (b) of Fig.4.
At the low temperature ($\beta =5.0$), $\sigma$ vanishes at odd  $N$ (except $N=1$), whereas it has peaks at even $N$'s. As the temperature increases, the dips and peaks at small and intermediate average numbers $N$ merge into a smooth and increasing curve but remain visible at sufficiently large values of $N$.

\section{Summary and concluding remarks}
In summary, we investigated the work distribution of many
non-interacting fermionic particles driven by an AB flux in
a non-simply connected geometry. In the single-particle case the
work distribution at high-temperatures, namely, in the classical
regime, is given by a Gaussian distribution, yielding $\langle
e^{-\beta w}\rangle$=1, indeed confirming that the quantum flux
leaves the free energy unchanged. By contrast, the distribution in
the quantum regime is found to be multimodal caused by particle
excitations. In particular, in order to deal with a many-particle system, we have
generalized the expression for the characteristic function of work to quantum systems that initially are in a grand canonical state. We proved that the difference of the grand potentials of a hypothetical grand canonical equilibrium system with the initial temperature and chemical potential at the at the final parameter values and of the actual initial system enters a generalized Jarzynski equality.
Although an energy measurement is an
experimentally challenging task, theoretical examination of work
in quantum many-particle systems {\it per se} is worthwhile for
the fundamental understanding of nonequilibrium characteristics.

{\it Acknowledgments.} This work was supported by the National Research Grant funded by the Korean Government~(NRF-2010-013-C00015), and the Volkswagen Foundation (project I/80424).


\begin{thebibliography}{99}
\bibitem{ab}
Y. Aharonov and D. Bohm, Phys. Rev. {\bf 115}, 485 (1959).

\bibitem{ac}
Y. Aharonov and A. Casher, Phys. Rev. Lett. {\bf 53}, 319 (1984).

\bibitem{ac2}
C.~R. Hagen, Phys. Rev. Lett. {\bf 64}, 2347 (1990).

\bibitem{ac3} A.~V. Balatsky
and B.~L. Altshuler, Phys. Rev. Lett. {\bf 70}, 1678 (1993).

\bibitem{dualac}
M. Wilkens, Phys. Rev. Lett. {\bf 72}, 5 (1994).

\bibitem{Jarzynski}
C.~Jarzynski, Phys. Rev. Lett. {\bf 78}, 2690 (1997).

\bibitem{Jarzynski2}
C.~Jarzynski, C.~R. Phys. {\bf 8}, 495 (2007).

\bibitem{classical1}
F.~Douarche, S. Ciliberto, A. Petrosyan, and I. Rabbiosi,
Europhys. Lett. {\bf 70}, 593 (2005).

\bibitem{classical2}
C. Bustamante, J. Liphardt and F. Ritort, Phys. Today {\bf 58}~(7) 43 (2005).

\bibitem{classical3}
V.Blickle, T. Speck, L. Helden, U. Seifert, and C. Bechinger, Phys. Rev. Lett. {\bf 96}, 070603 (2006).

\bibitem{Pradhan}
P. Pradhan, Phys. Rev. E {\bf 81}, 021122 (2010).

\bibitem{quantum1}
H. Tasaki, arXiv:cond-mat/0009244.

\bibitem{quantum2}
S. Mukamel, Phys. Rev. Lett. {\bf 90}, 170604 (2003).

\bibitem{quantum2b} M. Esposito and S. Mukamel, Phys. Rev.
E {\bf 73}, 046129 (2006).

\bibitem{quantum3}
W. De Roeck and C. Maes, Phys. Rev. E {\bf 69}, 026115 (2004).

\bibitem{TLH}
P. Talkner, E. Lutz, and P. H\"{a}nggi, Phys. Rev. E {\bf 75}, 050102(R) (2007).

\bibitem{TH}
P. Talkner and P. H\"{a}nggi, J. Phys. A {\bf 40}, F569 (2007).

\bibitem{THM}
P. Talkner, P. H\"{a}nggi, and M. Morillo, Phys. Rev. E {\bf 77}, 051131 (2008).

\bibitem{Talkner08PRE78}
P. Talkner, P. S. Burada, and P. H\"anggi,
Phys. Rev. E {\bf 78}, 011115 (2008).

\bibitem{Talkner09PRE79}
P. Talkner, P. S. Burada, and P. H\"anggi,
Phys. Rev. E. {\bf 79}, 039902 (2009).

\bibitem{DL}
S. Deffner and E. Lutz, Phys. Rev. E {\bf 77}, 021128 (2008).

\bibitem{ftopen}
M. Campisi, P. Talkner, and P. H\"{a}nggi, Phys. Rev. Lett. {\bf 102}, 210401 (2009).


\bibitem{EHM}
M. Esposito, U. Harbola, S. Mukamel, Rev. Mod. Phys. {\bf 81}, 1665 (2009).

\bibitem{CHT}
M. Campisi, P. H\"anggi, P. Talkner, arXiv:1012.2268.

\bibitem{pc}
R. Landauer and M. B\"{u}ttiker, Phys. Rev. Lett. {\bf 54}, 2049
(1985).

\bibitem{pc2}
V. Chandrasekhar, R.~A. Webb, M.~J. Brady, M.B. Ketchen, W.~J. Gallagher, and A. Kleinsasser, Phys. Rev. Lett. {\bf 67},
3578 (1991).

\bibitem{pc3}
M. Y. Choi, Phys. Rev. Lett. {\bf 71}, 2987 (1993).

\bibitem{loss}
D. Loss and P. Goldbart, Phys. Rev. B {\bf 43}, 13762 (1991).

\bibitem{pcexp}
A.~C. Bleszynski-Jayich, W.~E. Shanks, B. Peaudecerf, E. Ginossar, F. von Oppen, L. Glazman, and J.~G.~E Harris,
Science {\bf 326}, 272 (2009).

\bibitem{note}
In order to make the $\delta$-peaks visible we introduce a small positive number $\epsilon$ and plotted
$P(w)=\int_{-\infty}^{\infty} du e^{-i[u-i\epsilon
sgn(u)]w}G(u)$ with $sgn(x)=1$ if $x>0$, and $sgn(x)=-1$,
otherwise.

\bibitem{TCH} P. Talkner, M. Campisi, P. H\"anggi, J. Stat. Mech.: Theory Exp. P02025 (2009).

\bibitem{seifert}
T. Schmiedle and U. Seifert, J. Chem. Phys. {\bf 126}, 044101 (2007).

\bibitem{SU}
K. Saito, Y. Utsumi, Phys. Rev B {\bf 78}, 115429 (2008).

\bibitem{qmnumber}
D. Andrieux, P. Gaspard, T. Monnai, and S. Tasaki, New J. Phys.
{\bf 11}, 043014 (2009).

\bibitem{CTH} M. Campisi, P. Talkner, P. H\"anggi, Phys. Rev. Lett. {\bf 105}, 140601 (2010).


\end{thebibliography}
\end{document}